\documentclass[12pt]{statsoc}
\usepackage[margin=1in]{geometry}
\usepackage{amsmath,graphicx,amssymb,multicol,pdflscape,subfigure,captcont}
\newcommand{\br}{\boldsymbol{r}}
\newcommand{\E}{\mathbb{E}}
\newcommand{\bPsi}{\boldsymbol\Psi}

\title[Latent Space Models for Ranked Dynamic Networks]{Analysis of the Formation of the Structure of Social Networks using Latent Space Models for Ranked Dynamic Networks}
\author[]{Daniel K. Sewell}
\address{University of Illinois,
{Urbana-Champaign},
USA}
\email{dsewell2@illinois.edu}
\author[D. K. Sewell and Y. Chen]{Yuguo Chen}
\address{University of Illinois,
{Urbana-Champaign},
USA}

\begin{document}
\begin{abstract}
The formation of social networks and the evolution of their structures have been of interest to researchers for many decades.  We wish to answer questions about network stability, group formation and popularity effects.  We propose a latent space model for ranked dynamic networks that can be used to intuitively frame and answer these questions.  The well known data collected by Newcomb in the 1950's is very well suited to analyze the formation of a social network.  We applied our model to this data in order to investigate the network stability, what groupings emerge and when they emerge, and how individual popularity is associated with individual stability.
\end{abstract}

\vfill

{{\it Address for correspondence}:
Yuguo Chen,
Department of Statistics, University of Illinois at
Urbana-Champaign, 725 S. Wright Street, Champaign, IL 61820, USA.\\
E-mail: yuguo@illinois.edu
}

\section{Introduction}

The formation and evolution of interpersonal relationships are highly studied in the social sciences.  These interpersonal relationships can most easily be thought of in the context of a social network in which we observe how a certain number of actors interact.  By analyzing such a network over time, one can hope to quantify the construction and stabilization of the network and its structures.  In 1954 T. Newcomb began an observational study using a college fraternity for this purpose, and a very large number of researchers have relied on this study to help understand how social networks form and stabilize.  This fraternity data set gives social scientists the unique opportunity to study the evolution and formation of the structure of social networks from a nonexistent state to a stabilized form.  The overall goal of the original study was to ``improve our understanding of the development of stable interpersonal relationships" \citep{newcomb1961acquaintance}.

Some authors have used Newcomb's fraternity data as a example with which to illustrate new methodology, e.g., \cite{snijders1996stochastic} states ``our treatment of Newcomb's fraternity data in this paper is not more than an example \ldots"  Other authors have analyzed this data set in more depth, utilizing it for its worth in helping to understand social networks and how they form.  A notable example includes \cite{doreian1996brief}, who studied this data to determine how reciprocity, transitivity and group balance, as determined by how well the actors can be partitioned, vary over time.  Another such example can be found in \cite{krackhardt2007heider}, where the authors used this data to determine the significance of Heiderian triads and Simmelian triads.

We have three questions in particular we attempt to answer in this paper regarding Newcomb's fraternity data.  First, does the network stabilize, and if so, when does this happen?  Second, how do subgroups form and stabilize?  That is, do some or all of the actors naturally fall into a small number of groups, and if so when do these groups form?  Third, is there a relationship between the popularity of an individual and the social position of that individual?  We desire a unifying framework with which we can answer all three of these questions.

There exists vast literature on modeling static networks, and many models for these static networks have been extended to account for longitudinal, or dynamic, networks.  For example, the exponential random graph model (ERGM) was extended by \cite{hanneke2010discrete}, the wide class of blockmodels for static networks was extended by \cite{xing2010state}, and the latent space model derived by \cite{hoff2002latent} was extended by \cite{sarkar2005dynamic} and Sewell and Chen (2014).  Other models have been developed specifically for dynamic network data.  For example, continuous Markov processes have been used early on by \cite{holland1977dynamic} and more recently in the development of the stochastic actor oriented models \citep[see, e.g., ][]{snijders1996stochastic,snijders2010maximum}, and \cite{krivitsky2014separable} has done further work on the discrete-time model of \cite{hanneke2010discrete}.  However, in most cases it is not obvious how to further extend these models for weighted edges (though it must be said that Snijders (1996) has applied his model to Newcomb's fraternity, but only as a toy example using ranks in a somewhat ad hoc manner).  A particular challenge is appropriately modeling the type of ranked network data which we find in Newcomb's fraternity data, where each actor ranks each other actor from most to least favored.  Some work has been done for this type of data by \cite{gormley2007latent}, who combined the latent space model by \cite{hoff2002latent} with the Plackett-Luce model for ranked data in order to model a static bipartite network.  Other work in this was done by \cite{krivitsky2012rank}, who extended the ERGM for such ranked network data.

Most of the past analyses of Newcomb's fraternity data, however, have needed to simplify the data to complete their analyses.  For example, \cite{breiger1975algorithm} only considered the top two and the bottom two rankings for each individual during the final week of the study; \cite{arabie1978constructing} similarly used the top two and the bottom three rankings for each individual during the final week.  \cite{wasserman1980analyzing} tried using the top four and the top eight rankings to transform the ranked network into a binary network; as may be expected, Wasserman found that the network structure is affected by the binary cutoff.  \cite{doreian1996brief} used in parts of their analysis only the top four rankings, and in other parts used the top four and the bottom three.  More recently, \cite{moody2005dynamic} used only the top four rankings, and \cite{krackhardt2007heider} used the top eight.
Using the methods of \cite{SewellChen14}, we analyzed the fraternity data using the top four as edges and using the top eight as edges.  The resulting two visualizations of the network differed considerably from each other, and both gave different visualizations than that obtained in our final model (see Figures \ref{fratLatPos1} and \ref{fratLatPos2} for the visualizations obtained from our proposed model).  This suggests that, rather than selecting some arbitrary cutoff value we ought to try to model the full data.  For more on this topic see \cite{thomas2011valued}.  One last note is that a common theme among the analyses of Newcomb's fraternity data is that the network inference is based on ad hoc measures.  While these methods can still be useful, it is clearly more preferable to have a more rigorous model and estimation method which can elicit more confidence in the estimates and quantify uncertainties.

In this paper, we propose a latent space model for ranked dynamic network data.  Our approach avoids deciding on an arbitrary cutoff for binarizing the network by appropriately modeling the rank data.  Our approach also models the temporal dependence structure involved in observing the network over time.  Using a latent space approach to dynamic network data allows us to obtain an intuitive visualization of the network and its evolution, giving us a better understanding of the network and allowing us to make qualitative inference.  Further, by using a latent space approach we have an intuitive way to think about the network stability by linking network stability with how stable the actors' social positions are.  That is, if the network is not stable, then the actors' social positions ought to vary considerably from one time point to the next; however, as the network stabilizes, the social positions in turn ought to stabilize and vary less over time.  Our model allows us to measure the statistical precision of the movements of these social positions over time.  Our proposed model and estimation method allows us to quantify the uncertainty of the latent positions.  This uncertainty allows us to analyze group structure emergence.  Finally, our model also incorporates popularity measures, thereby capturing some of the local structure.  These popularity measures, together with the latent positions, can tell us about the relationship between individual popularity and individual stability.

While the main purpose of this paper is to analyze Newcomb's fraternity data, developing tools for rank-order network data is important in its own right.  Ranked networks should inherently contain more information than binary networks.  While it is true that rank-order network data is much rarer than binary data, it seems likely that this is due to a lack of analytical tools available.  This work adds to the current analytical toolbox, thereby encouraging researchers to collect and analyze ranked network data.

The remainder of the paper is as follows.  Section \ref{Data} describes the data; Section \ref{Model} describes the proposed model; Section \ref{Estimation} gives the estimation algorithm;
Section \ref{Results} gives the results of analyzing Newcomb's fraternity data;  Section \ref{Discussion} provides a brief discussion.

\section{Newcomb's Fraternity Data}
\label{Data}
In 1955, seventeen unacquainted students took part in a semester long study at the University of Michigan.  These students were selected in such a way that they were all unknown to each other before the study began.  Thus the data on a social network would be collected over time, beginning in its most nascent state and observed as the network evolves and stabilizes to its final form.  This purposeful capturing of the emergence of a social network is why this data is still of such interest nearly six decades later.  For fifteen out of sixteen weeks in the semester (no responses were recorded for week 9), each student would then rank the sixteen other students from most to least favored.  See \cite{newcomb1961acquaintance} Chapter 2 for details on the selection of the students and the data acquisition process.

Thus the data come in the form of a sequence of adjacency matrices $Y_t$  for $t=1,\ldots,15$.  For each time point $t$, the $i^{th}$ row of $Y_t$, denoted as ${\bf y}_{it}=(y_{i1t},y_{i2t},\ldots,y_{int})$, is a permutation of $\{1,2,\ldots,n-1\}$ with a 0 inserted into the $i^{th}$ position.  The rankings go, in order of most favored to least favored, from 1 to $n-1$.

\section{Models}
\label{Model}
We first describe our proposed model in Section \ref{LSMRDN}.  This methodology allowed us to gain insight into the stability of the network, as well as to investigate subgroup formation and the relationship between individual stability and individual popularity.  In Section \ref{Multilinear} we review the model derived by \cite{hoff2011hierarchical}.  We used this model to investigate the stability of the fraternity network over time, and while this did not detect all of the stability patterns that our proposed approach detected, it corroborated our main results on the timing of the network stability.

\subsection{Latent Space Hierarchical Model for Ranked Dynamic Networks}
\label{LSMRDN}
Due to the lack of existing methods for our context, we develop a latent space model for handling ranked longitudinal network data with which to answer our research questions.  We assume here that each actor exists within a latent space which can be interpreted as a characteristic space, or a social space.  This is the underlying concept of the latent space: a smaller distance between two actors within this space corresponds to a larger probability of receiving a favorable ranking.  Therefore if two nodes are far apart in the latent space we would expect them to rank each other unfavorably, whereas if two nodes are close together we would expect them to view each other quite favorably.

First is some general notation to be used throughout, following that of \cite{SewellChen14}.  Assume we have a set of actors ${\cal N}$ and a set of edges ${\cal E}$; let $n=|{\cal N}|$ be the fixed number of actors and $T$ the total number of time points at which the network is observed.  Often it will be more convenient to work with the ordering of ${\bf y}_{it}$ rather than ${\bf y}_{it}$ itself.  We will let $\boldsymbol\omega_{it}=(\omega_{i1t},\omega_{i2t},\ldots,\omega_{i(n-1)t})$ be the $(n-1)\times1$ vector which is the ordering of the rank vector ${\bf y}_{it}$ (e.g., if ${\bf y}_{1t}=(0,3,1,4,2)$ then $\boldsymbol\omega_{1t}=(3,5,2,4)$).  Let ${\bf X}_{it}\in\Re^p$ be the position vector of the $i^{th}$ actor at time $t$ within the $p$ dimensional latent space.  Let ${\cal X}_t$ be the matrix whose $i^{th}$ row is ${\bf X}_{it}$.  Finally, let $\bPsi$ be the vector of unknown parameters to be defined later.

We assume the actors' latent positions transition according to a Markov process, where the initial distribution is
\begin{equation}
\pi({\cal X}_1|\bPsi)= \prod_{i=1}^n N({\bf X}_{i1}|{\bf 0},I_p/\tau_0),
\label{transition1}
\end{equation}
and the transition equation is
\begin{equation}
\pi({\cal X}_t|{\cal X}_{t-1},\bPsi)=\prod_{i=1}^n N({\bf X}_{it}|{\bf X}_{i(t-1)},I_p/\tau_t) ,
\label{transition2}
\end{equation}
for $t=2,3,\ldots,T$, where $I_p$ is the $p\times p$ identity matrix, and $N({\bf x}|\boldsymbol{ \mu},\Sigma)$ denotes the multivariate normal probability density function with mean $\boldsymbol\mu$ and covariance matrix $\Sigma$ evaluated at ${\bf x}$.

The precision parameters $\tau_t$, $t=2,\ldots,T$, give us the information we need to evaluate the stability of the network.  A larger precision implies that the latent positions are moving less and therefore implies the actors' positions are more stabilized, whereas a smaller precision implies that the latent positions are moving more and therefore implies less stable social positions.  The network's stability at time $t$ ought to be in some sense smooth over time; that is, one would not expect the stability of the network at time $t$ to be drastically different from the stability at $t-1$ and $t+1$.  For this reason we further model the precision parameters $\tau_t$, $t\geq2$, as a random walk involving gamma distributed random variables.  Specifically we have for $t\geq2$ that
\begin{equation}
\tau_t=\tau_{t-1}\eta_t,
\label{tautCondPrior}
\end{equation}
where $\eta_t \stackrel{iid}{\sim} \Gamma(\theta,\theta)$, and $\Gamma(a,b)$ indicates a gamma distribution with shape parameter $a$ and rate parameter $b$.  This is equivalent to having the prior
\begin{equation}
\pi(\tau_2,\ldots,\tau_T)\stackrel{{\cal D}}{=} \prod_{t=2}^T \Gamma(\tau_t|\theta,\theta/\tau_{t-1}),
\label{GammaRandomWalk}
\end{equation}
where $\Gamma(x|a,b)$ is the gamma density function with shape $a$ and rate $b$ evaluated at $x$.  With this specification, $\tau_t$ conditional on $\tau_{t-1}$ has an expected value equal to $\tau_{t-1}$ and variance equal to $\tau_{t-1}^2/\theta$.  Note that $\tau_1$ is a hyperparameter that defines the mean of $\tau_2$ (and therefore the unconditional mean for any $\tau_t$, $t\geq2$).

The choice of $p$, the dimension of the latent space, is a topic that is beyond the scope of this paper.  As visualization of the network is a motivation for using the latent space approach to modeling networks, typically $p$ is set to two or three.  In our analysis we set $p=2$.

\begin{figure}[t]
\vspace{-3pc}
\hspace{-5pc}\includegraphics[scale=0.6]{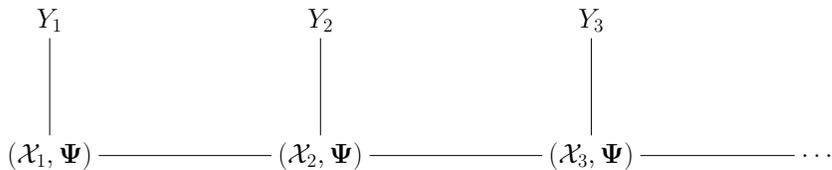} \vspace{-21pc}
\caption{Illustration of the dependence structure for the latent space model.  $Y_t$ is the observed graph, ${\cal X}_t$ is the unobserved latent actor positions, and $\bPsi$ is the vector of model parameters.}
\label{dep_struct}
\end{figure}

Many methods, such as the temporal exponential graph model by \cite{hanneke2010discrete} or the stochastic actor oriented models originated by \cite{snijders1996stochastic}, construct the dependence structure through modeling specific dependency structures; latent space approaches, such as our proposed model, assume that the dependency within the network has been induced by the latent variables.  Specifically, we assume that the observed networks at differing time points are conditionally independent given the latent positions, and that the observed network at time $t$ depends only on the latent space positions at time $t$.  Figure \ref{dep_struct} illustrates this dependence structure.  We also assume that, conditioning on $({\cal X}_t,\bPsi)$, ${\bf y}_{it}$ is independent of ${\bf y}_{i't}$, $i\neq i'$.

We now describe the likelihood component of the model that relates the distances between the latent positions and the observed network.  To this end we utilize the Plackett-Luce model for ranked data \citep[see][]{plackett1975analysis}.  The Plackett-Luce model can be thought of as drawing from a vase.  Every member of the set $\{1,\ldots,n\}\setminus \{i\}$ being ranked by $i$ has a particular proportion of the tickets with their name on it in the vase.  At time $t$, $i$ randomly draws a ticket and the name on the ticket determines who is ranked first, i.e., $\omega_{i1t}$.  For the second rank, $i$ draws until a new name is drawn and then ranks that name second, $\omega_{i2t}$.  This continues until all elements in the set are ranked.  Notice that the second rank is obtained according to the same probability distribution as if $i$ was deciding the first rank with the smaller set of $n-2$ elements, i.e., $\{1,\ldots,n\}\setminus\{i,\omega_{i1t}\}$.  In other words, $i$ ranks $j$ above $k$ with the same probability with and without $\ell$ included in the set to be ranked; this condition is called Luce's Choice Axiom.  It is reasonable to assume that this axiom holds;  if Newcomb had only asked a subset of the students living within the fraternity to rank each other, we would not expect the resulting network to look different than a subnetwork of the full data we actually have, where all the students are included in the network.  Using this framework we can write the distribution for ${\bf y}_{it}$ as a product of conditional probabilities given as
\begin{equation}
\mathbb{P}({\bf y}_{it}) = \mathbb{P}(\boldsymbol\omega_{it})
 = \prod_{j=1}^{n-1}\mathbb{P}(\omega_{ijt}|\omega_{i1t},\omega_{i2t},\ldots,\omega_{i(j-1)t})=\prod_{j=1}^{n-1}\frac{\nu_{i\omega_{ijt}t}}{\sum_{\ell=j}^{n-1}\nu_{i\omega_{i\ell t}t}},
\label{PL1}
\end{equation}
where, following the explanation given above, $\nu_{ijt}$ corresponds to the proportion of tickets with $j$'s name on it in $i$'s vase at time $t$.

As mentioned previously, we desire that the greater the distance between actor $i$ and actor $j$ the smaller the probability of each giving the other a favorable ranking.  Further, even within a common social circle there will still be more popular and less popular actors, and so it is important to capture this local structure in the model.  Therefore it is intuitive to model the $\nu_{ijt}$'s as functions of the latent positions and actor specific parameters.  The parameterization is chosen such that
\begin{equation}
\nu_{ijt}=r_j\exp(-d_{ijt}),
\label{PL_param}
\end{equation}
where $d_{ijt}=\|{\bf X}_{it}-{\bf X}_{jt}\|$ and $\br=(r_1, r_2, \ldots, r_n)$ is the vector of positive actor specific parameters constrained such that $\sum_{i=1}^nr_i=1$ for model identifiability.  These $r_i$'s can be interpreted as each actor's social reach, where a larger value implies a higher probability of receiving a favorable ranking from others.  Thus if an actor is generally well liked they will have a large $r_i$ value.  This parameterization is similar to that of \cite{gormley2007latent}, who applied the Plackett-Luce model to a bipartite network, though here we also incorporate the popularity measures into the likelihood.

From (\ref{PL1}) and (\ref{PL_param}) we have that the conditional likelihood of $(Y_1,Y_2,\ldots,Y_T)$ is
\begin{equation}
\mathbb{P}(Y_1,Y_2,\ldots,Y_T|{\cal X}_1,{\cal X}_2,\ldots,{\cal X}_T,\bPsi)=\prod_{t=1}^T\prod_{i=1}^n\prod_{j=1}^{n-1}\frac{r_{\omega_{ijt}}\exp(-d_{i\omega_{ijt}t}) }{\sum_{\ell=j}^{n-1}r_{\omega_{i\ell t}}\exp(-d_{i\omega_{i\ell t}t})},
\label{PL2}
\end{equation}
where $\bPsi=(\br,\tau_0,\tau_1,\tau_2,\ldots,\tau_T)$.

Further motivation for the parameterization in (\ref{PL_param}) is that we can consider the Thurstonian model interpretation of the Plackett-Luce model.  \cite{thurstone1927law} described the following model: For a vector of ranked data ${\bf y}=(y_1,y_2,\ldots,y_m)$, there is a vector of latent random variables ${\bf Z}=(Z_1,Z_2,\ldots,Z_m)$ and a vector of scalars $\boldsymbol\mu=(\mu_1,\mu_2,\ldots,\mu_m)$ such that $Z_j-\mu_j \stackrel{iid}{\sim} F$ for some continuous distribution function $F$.  Then $\mathbb{P}({\bf y})=\mathbb{P}(Z_{\omega_1}>Z_{\omega_2}>\cdots>Z_{\omega_m})$, where $\boldsymbol\omega=(\omega_1,\omega_2,\ldots,\omega_m)$ is the ordering of ${\bf y}$.  \cite{yellott1977relationship} showed that the Plackett-Luce model is equivalent to the Thurstone model if and only if $F$ is the Gumbel distribution.  They further showed that if $F$ is a Gumbel distribution with location parameter equal to zero and scale parameter equal to 1, then the relationship between the two models is that $\nu_j=\exp(\mu_j)$.  Coming back to our context, we let ${\bf Z}_{it}=(Z_{i1t},Z_{i2t},\ldots,Z_{i(n-1)t})$ be a vector of latent random variables which measure how actor $i$ regards the strength of his/her relationship with the other $n-1$ actors.  We define these measures such that
\begin{equation}
Z_{ijt}=\mu_{ijt}+\epsilon_{ijt}
\label{Zijt}
\end{equation}
where $\mu_{ijt}=\log(r_j)-d_{ijt}$, $\epsilon_{ijt}\stackrel{iid}{\sim}F=\mbox{Gumbel}(-\gamma_{EM},1)$, and $\gamma_{EM}$ is the Euler-Mascheroni constant ($\approx 0.5772$); the location shift is because a Gumbel(0,1) random variable has mean $\gamma_{EM}$ and thus by including the location shift we set the mean of $\epsilon_{ijt}$ to be zero.  Note also that the non-zero location parameter of $F$ does not change the relationship between the Thurstonian model and the Plackett-Luce model.  To see why this is so, it is necessary to recognize that the Plackett-Luce model is invariant to rescaling the $\nu_{ijt}$'s, and hence we can rescale by $\exp(-\gamma_{EM})$.  By the relationship mentioned above, we have that $\nu_{ijt}=\exp(\mu_{ijt}-\gamma_{EM})$, hence $Z_{ijt}-(\mu_{ijt}-\gamma_{EM})\stackrel{iid}{\sim}\mbox{Gumbel}(0,1)$, which is equivalent to (\ref{Zijt}).  This meets our intuition that the ranking of the $Z_{ijt}$'s should not be affected by a location shift of $F$.  The actual reason we desire this non-zero location parameter of $F$ is so that we have
\begin{equation}
\mathbb{E}(Z_{ijt}|{\cal X}_t,\br)=\log(r_j)-d_{ijt}.
\label{EZijt}
\end{equation}
Therefore the Plackett-Luce model in (\ref{PL2}) can be thought of as, for individual $i$ at time $t$, obtaining a set of variables $Z_{ijt}$, $j\neq i$, whose mean is determined by the social reach of the actor being ranked and by the social distance between the ranking actor and the ranked actor, which measures on a continuous scale the relationship between individual $i$ and the rest (as perceived by $i$).  Then the vector ${\bf y}_{it}$ is the ranking of the realizations $z_{ijt}$ of $Z_{ijt}$.

\subsection{Multilinear Model for Multiway Data}
\label{Multilinear}
\cite{hoff2011hierarchical} developed a latent space approach for analyzing multiway data, which he then demonstrated how to apply the model on dynamic network data.  In particular, he applied his model to a dynamic network whose edges $y_{ijt}$ consist of ranking the relationship on the constant set $\{-5,-4,\ldots,2\}$.  This type of ranked network is different than the fraternity data, where there is the added constraint on the response variables that the rows of the response array must be a permutation of $\{0,1,\ldots,n-1\}$.  In applying this model to the fraternity data set, we relax this extra constraint, thereby allowing the model to predict networks that violate the permutation constraint.  This can be thought of as another form of simplifying the network at some cost to the information contained therein, much like, and arguably to a much lesser degree than, the information lost associated with transforming the network from weighted to binary according to some arbitrary cutoff.  Hoff's model utilized an ordered probit model, which we now briefly describe within the context of the fraternity data.

Let $z_{ijt}$ be latent variables such that $y_{ijt}=\max\{k: z_{ijt}>c_k, k\in\{1,\ldots,n-1\}\}$, where the $c_k$'s are unknown cutoff points to be estimated.  These latent variables are assumed to be normally distributed whose mean can be written as the following factor model:
\begin{equation}
\E(z_{ijt})=\sum_{\ell=1}^p u_{i\ell}u_{j\ell}v_{t\ell}.
\end{equation}
The $p$ dimensional vectors ${\bf u}_i=(u_1,\ldots,u_p)$ are student specific vectors that can be equated to the latent positions ${\bf X}_{it}$ in the model of Section \ref{LSMRDN}, though instead of being time dependent, in Hoff's model the temporal aspect of the data is accounted for by the $p$ dimensional vectors ${\bf v}_t=(v_{t1},\ldots,v_{tp})$.  The ${\bf u}_i$'s can then be thought of as the time invariant latent positions of the students, and the ${\bf v}_t$'s can be thought of as stretching or compressing the $p$ axes to alter the closeness of the students at different time points.  There are no structural constraints placed on the ${\bf u}_i$'s and ${\bf v}_t$'s beyond the regularization that the Bayesian framework imposes via the prior distributions.  Note also that the closeness between the actors is not measured via Euclidean distance, as in Section \ref{LSMRDN}, but rather by the cosine of the angle between the two students, more akin to the dot product graph model \citep[see, e.g.,][]{young2007random}.  The dimension $p$, just as in our proposed approach, is assumed to be 2, though this is in actuality an unknown quantity.  Hoff suggested using the Deviance Information Criterion \citep[][]{spiegelhalter2002bayesian}, though determining the optimal $p$ could and should be a topic of future research.

The usefulness of this model within our context lies in the values of the ${\bf v}_t$'s.  These vectors give us a good sense as to the stability of the network, as conceptualized by how much the students' social positions are changing over time.  For example, if the network is completely stabilized over a set of time points ${\cal T}$ then the students' positions are static, and thus ${\bf v}_t={\bf v}_s$ for $s,t\in{\cal T}$.  If, on the other hand, the network is quite unstable, then we would expect to see these ${\bf v}_t$'s to vary considerably from week to week during the unstable time period.

Estimation for this model was performed by first running a Markov chain Monte Carlo (MCMC) algorithm to initialize the unknown quantities, and then applying an alternating least squares algorithm to obtain point estimates of the ${\bf u}_i$'s and ${\bf v}_t$'s.  Section \ref{Estimation} gives the details on the estimation procedure for our proposed model given in Section \ref{LSMRDN}.

\subsection{Pseudo-$R^2$}
In the context of linear regression, one can determine how well the model explains the data by using the $R^2$ or adjusted $R^2$ value.  For standard ranked data, there exist some measures that are approximately equivalents (see, e.g., Marden, 1995).
However, we cannot apply these measures to our context due to having each actor ranking a different set, i.e., each $i$ ranks the set $\{1,2,\ldots,n\}\setminus i$.  For the ordinal probit model, \cite{mckelvey1975statistical} devised a goodness of fit measure; \cite{veall1992pseudo} showed that McKelvey and Zavoina's pseudo $R^2$ is closest to the $R^2$ corresponding to the underlying continuous (latent) data.  We developed a pseudo-$R^2$ with a similar flavor by using the Thurstonian model specification outlined at the end of Section \ref{LSMRDN}.  Specifically, we note that
$$(z_{ijt}-\bar{z})^2=(\mu_{ijt}-\bar{z})^2+\epsilon_{ijt}^2+2\epsilon_{ijt}(\mu_{ijt}-\bar{z}),$$
where $\mu_{ijt}=\log(r_j)-d_{ijt}$ and $\bar{z}=1/(Tn(n-1))\sum_t\sum_{i\neq j} z_{ijt}.$   Since $\epsilon_{ijt}\stackrel{iid}{\sim}\mbox{Gumbel}(-\gamma_{EM},1)$, $\mathbb{E}(\epsilon_{ijt})=0$ and $Var(\epsilon_{ijt})=\pi^2/6$; thus we have that
 \begin{eqnarray}\nonumber
&&\sum_{t=1}^T\sum_{i\neq j}\mathbb{E}(\epsilon_{ijt}^2+2\epsilon_{ijt}(\mu_{ijt}-\bar{z}))\\ \nonumber
&=& \sum_{t=1}^T\sum_{i\neq j}\mathbb{E}(\epsilon_{ijt}^2) -2\sum_{t=1}^T\sum_{i\neq j}\mathbb{E}\left(\epsilon_{ijt}\frac{1}{Tn(n-1)}\sum_{t'=1}^T\sum_{i'\neq j'}(\mu_{i'j't'}+\epsilon_{i'j't'})\right)\\
&=&\frac{\pi^2}{6}(Tn(n-1)-2).
\end{eqnarray}
We can then, similarly to the method used by McKelvey and Zavoina, approximate the total sum of squares by
\begin{equation}
\sum_{t=1}^T\sum_{i\neq j}(z_{ijt}-\bar{z})^2\approx (\hat{\mu}_{ijt}-\hat{\bar{\mu}})^2 + \frac{\pi^2}{6}(Tn(n-1)-2),
\end{equation}
where $\hat{\mu}_{ijt} = \log(\hat{r}_j)-\hat{d}_{ijt}$, $\hat{\bar{\mu}}=1/(Tn(n-1))\sum_t\sum_{i\neq j}\hat{\mu}_{ijt}$, and the $\hat{}$ symbol over the model parameters implies the posterior mean estimate.  Therefore we define the pseudo $R^2$ to be
\begin{equation}
R^2=\frac{\sum_{t=1}^T\sum_{i\neq j}(\hat{\mu}_{ijt}-\hat{\bar{\mu}})^2}{\sum_{t'=1}^T\sum_{i'\neq j'}(\hat{\mu}_{i'j't'}-\hat{\bar{\mu}})^2 + \pi^2(Tn(n-1)-2)/6}.
\label{pseudoR2_rank}
\end{equation}
This $R^2$ value can be interpreted to be the approximate proportion of the variability of the underlying latent variables $z_{ijt}$ explained by the model; hence, all other things equal, we desire to have a higher $R^2$ value.

\section{Estimation}
\label{Estimation}
Estimation is done within a Bayesian framework; thus we desire to make inference based on the posterior distribution $\pi({\cal X}_1,\ldots,{\cal X}_T,\bPsi|Y_1,\ldots,Y_T)$.  The strategy is to find reasonable initial estimates of the latent positions and of the model parameters, and use these estimates to initialize a Metropolis-Hastings (MH) within Gibbs Markov chain Monte Carlo.  From the samples from the Markov chain we can then obtain posterior inference of the latent positions and of $\bPsi$.

To perform the Bayesian estimation, we first need priors on the model parameters.  We use the following:
\begin{align}
\pi(\br)&\stackrel{{\cal D}}{=}Dir(\alpha_1,\ldots,\alpha_n), & \\
\pi(\tau_0)&\stackrel{{\cal D}}{=} Exp(\lambda_0), & \\
\pi(\tau_1)&\stackrel{{\cal D}}{=}  \Gamma^{-1}(\lambda_1/2,1/2) , &\\
\pi(\theta)&\stackrel{{\cal D}}{=} LN(\mu,\sigma^2), &
\end{align}
where $Dir(\alpha_1,\ldots,\alpha_n)$ is the Dirichlet distribution, $Exp(a)$ is the exponential distribution with rate $a$, $\Gamma^{-1}(a/2,1/2)$ is the inverse gamma distribution with shape $a/2$ and scale $1/2$ (this is also the inverse-$\chi^2$ distribution with degrees of freedom $a$), and $LN(a,b)$ is the log-normal distribution with log-mean $a$ and log-variance $b$.  The Dirichlet is a natural prior for such constrained parameters as $\br$, the priors for $\tau_0$ and $\tau_1$ were chosen based on conjugacy, and the prior for $\theta$ was chosen to be able to put a flat prior on $\theta$ and also for ease of sampling.

\subsection{Initialization}
\label{Initialization}
In a complicated hierarchical model such as ours, it is difficult to know how to reasonably choose initial values of the Markov chain estimation algorithm or how to specify the hyperparameters of the prior distributions.  We attempt to address both these issues simultaneously via an approach which is similar in concept to empirical Bayes methods.  That is, we use the data to determine the initial values and the hyperparameters of the prior distributions.  The way in which we use the data is through a preliminary, and admittedly somewhat ad hoc, analysis of the data.  Therefore we make the priors flat and uninformative where possible, otherwise we use this preliminary analysis to determine the values of  the hyperparameters.  In so doing we naturally obtain initial values for the Markov chain estimation algorithm.

Since the social reaches should reflect the popularity of the individuals, we initialized the social reaches as
\begin{equation}
r_i^{(1)}=\frac{\sum_{t=1}^T \sum_{j=1}^n 2(n-y_{jit})}{n^2(n-1)T},
\label{init_weights_rank}
\end{equation}
where the superscript $(1)$ denotes the initial estimate.  These values account for how favorable student $i$ was with respect to all other students over all time points.  One could use $\br^{(1)}$ as the hyperparameters $\alpha_1,\ldots,\alpha_n$; in this case, however, we can make the prior distribution flat and uninformative by setting these hyperparameters all equal to one.  This also has the beneficial effect of reducing the computational complexity of the algorithm.

To find the initial latent positions we used classic multidimensional scaling (MDS) at each time point.  To implement this, we first needed a dissimilarity matrix for each time point.  We constructed this by setting
\begin{equation}
d_{ijt} \propto \frac{r_j^{(1)}}{n-y_{ijt}} + \frac{r_i^{(1)}}{n-y_{jit}}.
\label{distances}
\end{equation}
The logic behind this choice is that the more favorable $i$ and $j$ rank each other, the closer they ought to be in the latent space.  The latent social positions in our latent space model account for popularity, however, and so we use the initial values of the social reaches $\br^{(1)}$ to determine $d_{ijt}$.  The idea is that even if $i$ gives $j$ a favorable ranking, this may not imply that $i$ and $j$ are particularly close if $j$ has a large social reach.  If, however, $i$ gives $j$ a favorable ranking and $j$ has a very small social reach then this implies that $i$ and $j$ should be very close together in the latent social space.

With the $T$ dissimilarity matrices computed, we can then implement MDS to obtain initial latent positions. In many contexts \citep[e.g., see][]{SewellChen14} it would be more appropriate to initialize using the generalized multidimensional scaling derived by \cite{sarkar2005dynamic}, which implements MDS while accounting for the longitudinal aspect of the dissimilarity matrices.  However, this method implicitly assumes that $\tau_2=\tau_3=\cdots=\tau_T$, which we do not assume here; thus we have used a simpler MDS approach to initialize the latent positions, i.e., we use MDS on each of the $T$ dissimilarity matrices.  After each dissimilarity matrix has been used to embed the actors within a $p$-dimensional latent space, we used a Procrustes transformation to orient the latent positions at time $t$ as closely as possible to those at time $t-1$.  The Procrustes transformation finds a set of rotations, reflections and translations to minimize the difference between a given matrix and some target matrix \citep[see, e.g., ][]{borg2005modern}.  Lastly, we needed to know how to scale the latent positions.  To this end we maximized the likelihood using a simple line search to find
\[
c_0=\underset{c}{\mbox{argmax }} \pi(Y_1,\ldots,Y_T|c{\cal X}_1^*,\ldots,c{\cal X}_T^*,\br^{(1)}),
\]
and then we set ${\cal X}_t^{(1)}=c_0{\cal X}_t^*$ for $t=1,\ldots,T$, where ${\cal X}_t^*$ is the $t^{th}$ latent positions found by using MDS.

The prior mean of $\tau_0$ and the initial estimate $\tau_0^{(1)}$ was computed as
\begin{equation}
\left[\frac{1}{np}\sum_{i=1}^n\|{\bf X}_{i1}^{(1)}\|^2\right]^{-1}.
\label{tau0Init}
\end{equation}
We then set $\lambda_0=1/\tau_0^{(1)}$, thereby matching the prior expected value of $\tau_0$ to $\tau_0^{(1)}$.  Similarly, for $t\geq 2$, $\tau_t^{(1)}$ was computed as
\begin{equation}
\left[\frac{1}{np}\sum_{i=1}^n\|{\bf X}_{it}^{(1)}-{\bf X}_{i(t-1)}^{(1)}\|^2\right]^{-1}.
\label{tautInit}
\end{equation}
We set $\tau_1^{(1)}$ to equal $\tau_2^{(1)}$.  Matching the prior expected value of $\tau_1$ to equal $\tau_1^{(1)}$ implies setting $\lambda_1=2+1/\tau_1^{(1)}$.  Looking at (\ref{tautCondPrior}), we see that the variance of $\eta_t$ $(=\tau_t/\tau_{t-1})$ equals $1/\theta$.  Therefore we can set the initial estimate $\theta^{(1)}$ equal to the inverse of the sample variance of $\{\tau_t^{(1)}/\tau_{t-1}^{(1)},t\geq2\}$.  We then set $\mu=\log(\theta^{(1)})$ and set $\sigma^2$ to be some large value, thereby making the prior flat.

We checked the sensitivity to this initialization scheme on our analysis of the fraternity data.  Without getting into the details, which are given in the Supplementary Materials, we checked this sensitivity by choosing two alternative methods of initialization, each of which reflects some incorrect concept behind the latent space model (a misinterpretation of the latent positions and an assumption of constant network stability over time).  In neither case did the conclusions based on the samples from the posterior, which will be discussed in Section \ref{Results}, change.

\subsection{Posterior Sampling}
\label{PosteriorSampling}
To sample from the posterior distribution, we use a MH within Gibbs sampling scheme.  For this algorithm we need the full conditional distributions.  For the latent positions these are given as
\begin{align}\nonumber
&\pi({\bf X}_{it}|\cdot)& \\
&\propto \left\{\begin{array}{l l}
\pi(Y_1|{\cal X}_1,\bPsi)N({\bf X}_{i1}|{\bf 0},I_p/\tau_0)N({\bf X}_{i2}|{\bf X}_{i1},I_p/\tau_2) & \mbox{if $t=1$} \\
\pi(Y_t|{\cal X}_t,\bPsi)N({\bf X}_{it}|{\bf X}_{i(t-1)},I_p/\tau_t)N({\bf X}_{i(t+1)}|{\bf X}_{it},I_p/\tau_{t+1}) & \mbox{if $2\leq t<T$} \\
\pi(Y_T|{\cal X}_T,\bPsi)N({\bf X}_{iT}|{\bf X}_{i(T-1)},I_p/\tau_T) & \mbox{if $t=T$,}
\end{array}\right. & \label{XitCond}
\end{align}
and for the parameters are given as
\begin{align}
\pi(\br|\cdot)& \propto \pi(Y_1,\ldots,Y_T|{\cal X}_1,\ldots,{\cal X}_T,\bPsi) & \label{rCond} \\
\pi(\tau_2,\ldots,\tau_T|\cdot)&=\prod_{t=2}^T \Gamma\Big(\tau_t|\theta+\frac{np}{2},\frac{\theta}{\tau_{t-1}}+\frac{1}{2}\sum_{i=1}^n \|{\bf X}_{it}-{\bf X}_{i(t-1)}\|^2\Big) & \label{tautCond} \\
\pi(\tau_0|\cdot)&\stackrel{{\cal D}}{=} \Gamma\Big(1+\frac{np}{2},\lambda_0+\frac{1}{2}\sum_{i=1}^n\|{\bf X}_{i1}\|^2\Big) & \label{tau0Cond} \\
\pi(\tau_1|\cdot)&\stackrel{{\cal D}}{=} \Gamma^{-1}\Big( \frac{\lambda_1}{2} +\theta,\frac{1}{2}+\theta\tau_2\Big) & \label{tau1Cond} \\
\pi(\theta|\cdot)& \propto\left[ \prod_{t=2}^T \Gamma(\tau_t|\theta,\theta/\tau_{t-1})\right]\cdot LN(\theta|\mu,\sigma^2). & \label{thetaCond}
\end{align}

The algorithm is
\begin{description}
\item[0.] Set the initial values of the latent positions and parameters as given in Section \ref{Initialization}.\vspace{-1pc}
\item[1.] For $t=1,2,\ldots,T$ and for $i=1,2,\ldots,n$, draw ${\bf X}_{it}$ from (\ref{XitCond}) via MH. \vspace{-2.5pc}
\item[2.] Draw $\tau_0$ from (\ref{tau0Cond}).\vspace{-1pc}
\item[3.] Draw $\tau_1$ from (\ref{tau1Cond}).\vspace{-1pc}
\item[4.] For $t=2,\ldots,T$, draw $\tau_t$ from its conditional distribution in (\ref{tautCond}).\vspace{-1pc}
\item[5.] Draw $\theta$ from (\ref{thetaCond}) via MH.\vspace{-1pc}
\item[6.] Draw $\br$ from (\ref{rCond}) via MH.\vspace{-1pc}
\item[]Repeat steps 1-6.\vspace{0pc}
\end{description}
Regarding the proposal distributions, ${\bf X}_{it}$, $\beta_{IN}$, and $\beta_{OUT}$ can come from a symmetric proposal (e.g., normal random walk).  Because of the constraint on $\boldsymbol{r}$, a Dirichlet proposal is suggested for the radii, which also will be an asymmetric proposal.  Suggested parameters for this Dirichlet proposal are $\kappa\boldsymbol{r}^{curr}$, where $\boldsymbol{r}^{curr}$ are the current values for $\boldsymbol{r}$ and $\kappa$ is some large value (e.g., we set $\kappa=10,000$).

One final note is that, as is the case for any such latent space model, the posterior is invariant under rotations, reflections and translations of the latent positions ${\cal X}_1,{\cal X}_2, \ldots,{\cal X}_T$.  Hence after each iteration of steps 1-6, a Procrustean transformation will be performed on the $n$ trajectories; that is, the transformation is performed on the $nT\times p$ matrix $({\cal X}_1',{\cal X}_2',\ldots,{\cal X}_T')'$.  In our context, the target matrix is chosen to be constructed from the first MCMC draw of the latent positions after the burn-in.  In so doing we find a rotation matrix $A$ such that for any $i$ and $t$, ${\bf X}_{it}^{(\ell)}=A'{\bf X}_{it}^*$, where ${\bf X}_{it}^{(\ell)}$ is the stored latent positions for the $\ell^{th}$ iteration and ${\bf X}_{it}^*$ is the newly drawn latent positions.

\section{Results}
\label{Results}

\begin{figure}[!h]
\subfigure[$\theta$]{
\includegraphics[width=0.3\linewidth]{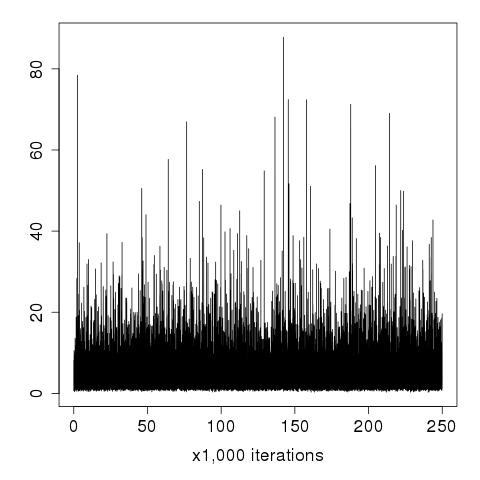}
}
\subfigure[$\tau_0$]{
\includegraphics[width=0.3\linewidth]{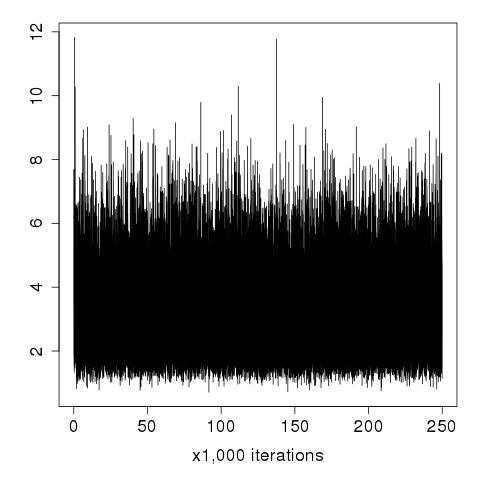}
}
\subfigure[$\tau_1$]{
\includegraphics[width=0.3\linewidth]{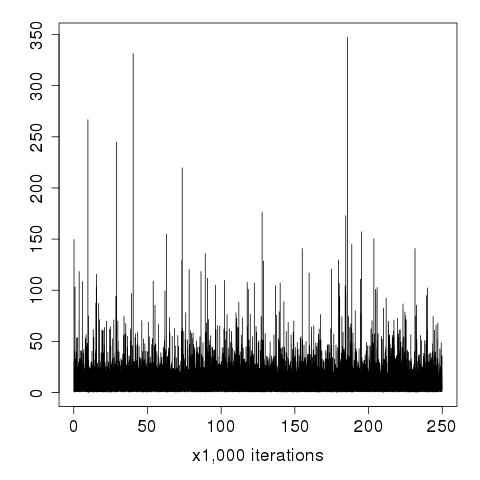}
} \\
\subfigure[$\tau_2$]{
\includegraphics[width=0.3\linewidth]{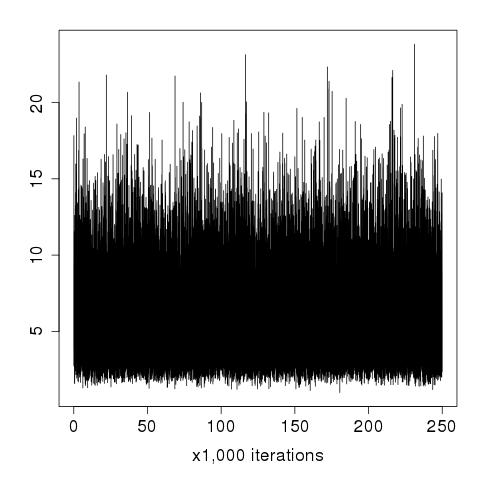}
}
\subfigure[$\tau_9$]{
\includegraphics[width=0.3\linewidth]{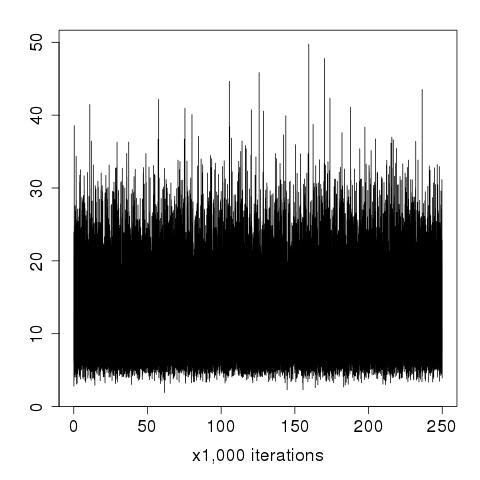}
}
\subfigure[$\tau_{15}$]{
\includegraphics[width=0.3\linewidth]{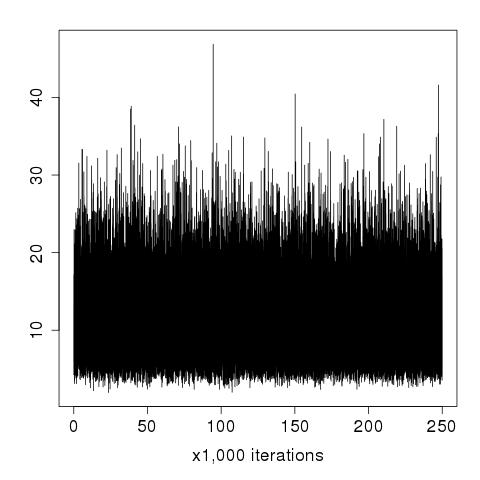}
}
\caption{Trace plots for select parameters corresponding to the analysis of Newcomb's fraternity data.}
\label{tracePlots}
\end{figure}

We applied our method to Newcomb's (1961) fraternity data .  We let the MCMC algorithm run for 250,000 iterations, including a burn in period of 50,000 iterations.  Figure \ref{tracePlots} gives the trace plots for selected parameters, namely $\theta$ and $\tau_t$ for $t=0,1,2,9,15$.  From this we see that the MCMC algorithm converges.  The hyperparameters $\alpha_1,\ldots,\alpha_{17}$ were all set to 1, $\sigma$, the log standard deviation of $\pi(\theta)$, was set to equal 5, and all other hyperparameters were chosen as described in Section \ref{Estimation}.

\begin{figure}
\centering
\includegraphics[width=5.5in]{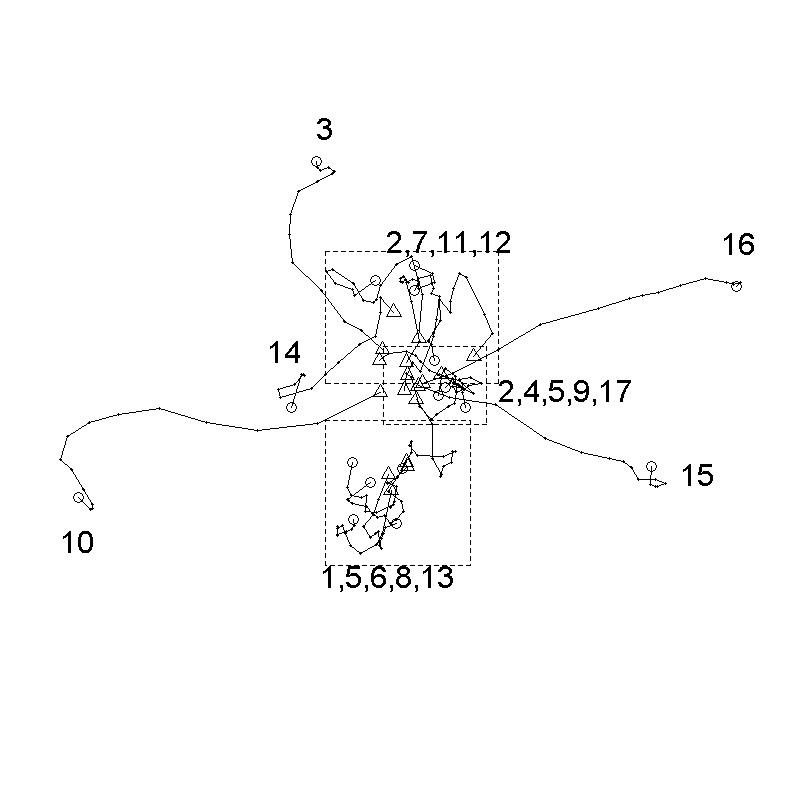}
\caption{Posterior means of the latent positions of the students in Newcomb's fraternity study.  Triangles indicate the beginning of the trajectory (week 1) and circles indicate the end of the trajectory (week 15).  When students' trajectories are obfuscated by each other, the students forming the group is given, rather than labeling each individual trajectory.}
\label{fratLatPos1}
\end{figure}

\begin{figure}
\centering
\subfigure[]{
\includegraphics[width=0.52\textwidth]{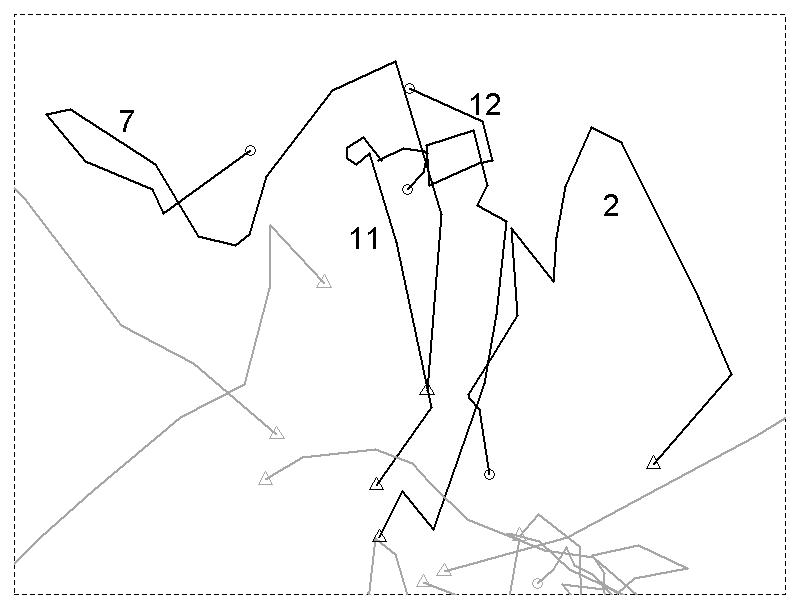}
}\\
\subfigure[]{
\includegraphics[width=0.52\textwidth]{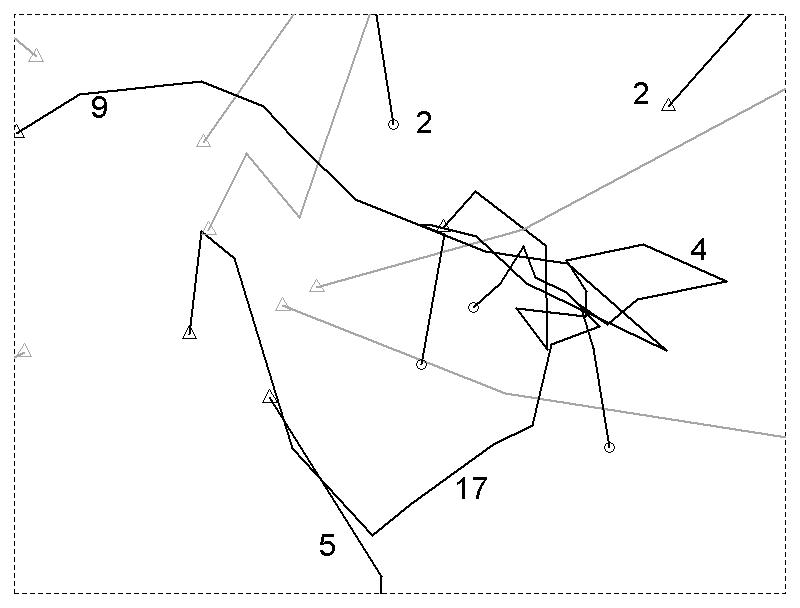}
}\\
\subfigure[]{
\includegraphics[width=0.52\textwidth]{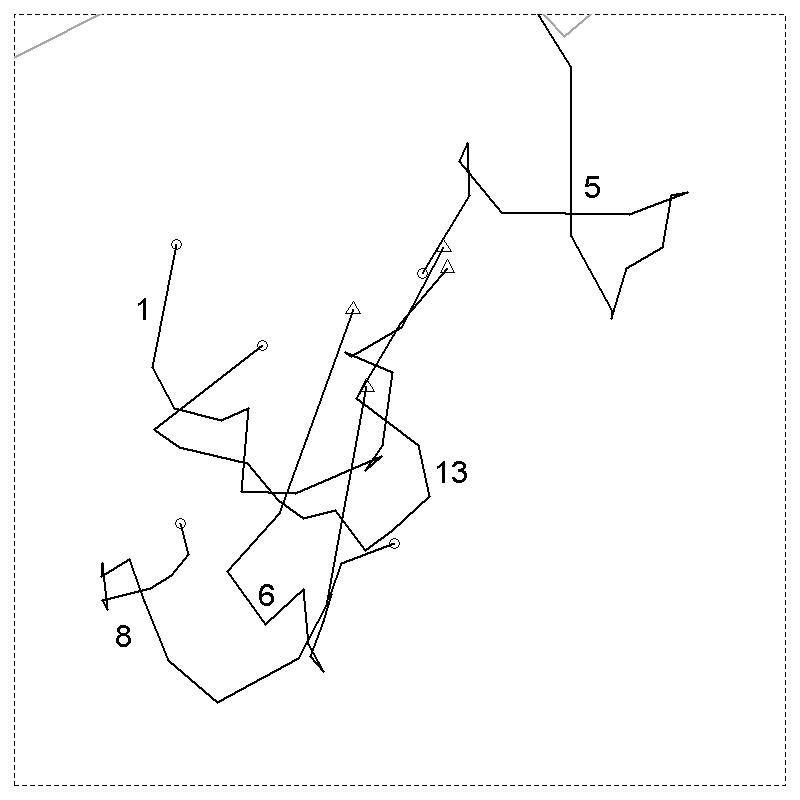}
}
\caption{Latent positions of the students; (a)--(c) zoom in on the top, central and bottom dashed boxes respectively of Figure \ref{fratLatPos1}, where the obfuscated student trajectories, in dark, are labeled.}
\label{fratLatPos2}
\end{figure}

The pseudo-$R^2$ value was 0.622 (this was equal up to three decimal places of the mean pseudo-$R^2$ values obtained from analyzing 20 data sets simulated from the model of Section \ref{LSMRDN} whose parameters were set to be equal to those learned from this data set; see the Supplementary Materials for details on the simulation study).  As this value approximates the amount of the variation in the underlying process explained by our model, we get some sense as to the noisiness of the data.  Our model has explained more than half of the variation of the latent process, though there is still some inherent unexplained noise in the network data.  Figures \ref{fratLatPos1} and \ref{fratLatPos2} give the posterior means of the latent trajectories of the 17 students through the 15 weeks of the study; also included in the Supplementary Materials is an mp4 video file showing the evolution of the network.  From this we get a better understanding of what the network looks like, what groupings exist, and which actors find their social positions early and which find their social positions late. The details are given in the following sections.


\subsection{Network Stability}
\cite{newcomb1961acquaintance} and \cite{nakao1993longitudinal} both measured the stability of the network by comparing each individual's rankings from week to week.  Newcomb claimed that the stability sharply increases in the first three weeks, and the network is essentially stable after this point.  Nakao and Romney claimed that the network is stable after week five.  Much more recently, \cite{krivitsky2012rank} extended the exponential random graph for ranked network data.  Krivitsky and Butts used this model to analyze Newcomb's fraternity data, determining the stability of the network through ranking inconsistencies, showing that according to this measure the stability of the network increases over time with a decrease at week 15.  We wish to use our model to conduct a formal analysis, giving quantitative answers to how the stability of the network evolves.  In so doing we verify the general trends discovered earlier, as well as discovering a new pattern in the stability of the network.

\begin{figure}
\centering
\includegraphics[scale=0.35]{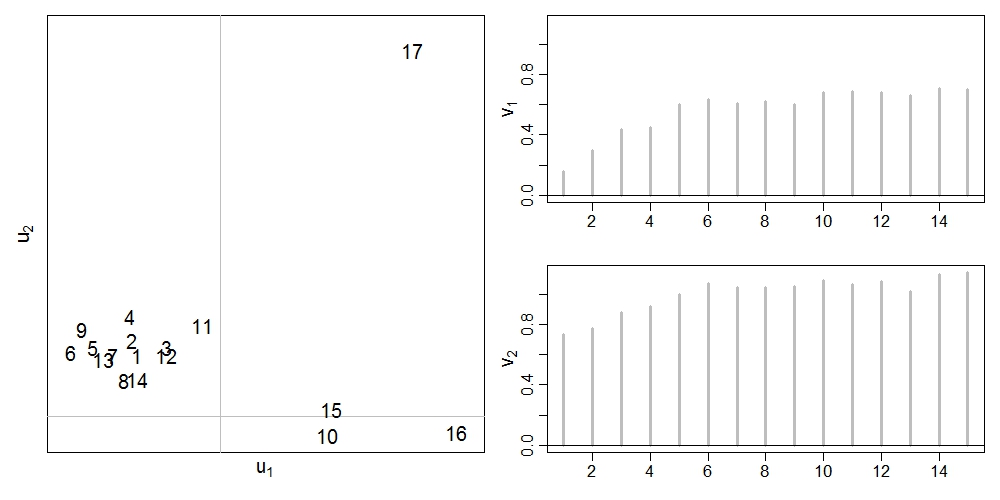}
\caption{Application of the multilinear model to the fraternity data.  The latent positions of the actors are given by ${\bf u}_1$ and ${\bf u}_2$, and the time effects are given by ${\bf v}_1$ and ${\bf v}_2$.}
\label{hoff}
\end{figure}

In a latent space approach to modeling the network, network stability is considered to be how constant the actor's social positions become.  Before applying our model from Section \ref{LSMRDN} for ranked dynamic networks, we first use Hoff's multilinear model to obtain a visualization of the evolution of the stability of the network.  Figure \ref{hoff} gives the resulting figures from the analysis.  Keep in mind that the interpretation of the latent positions from Hoff's model is different than that of the latent positions from our proposed method, in that a smaller angle, not a smaller distance, between the actors increases the probability of a favorable ranking.  The plots of ${\bf v}_1$ and ${\bf v}_2$ give the scalar time effects (which stretch the $g^{th}$ axis at time $t$ if $v_{tg}>1$ or contract if $v_{tg}<1$, $g=1,2$) for each of the two dimensions in the latent space.  It is these two plots which indicate how much the latent positions are moving over time.  During the first six weeks we see from Figure \ref{hoff} that the axes are being scaled  by different (increasing) factors, whereas from week six to the end of the study the axes are being scaled by a nearly constant factor.  This implies that for the first six weeks the latent positions are varying and thus the network is not stabilized, but after week six the latent positions are mostly static and hence the network is stable.
This result implies that both Newcomb and Nakao and Romney underestimated the time at which the network stabilised.

We next apply our proposed model for ranked dynamic networks to obtain more quantitative results on the evolution of the network stability. Again, in using a latent space approach to modeling the network we consider the network stability to be equivalent to the stability of the actors' social positions. While the stability of the social positions is an intuitive way of measuring network stability, we can understand even better how the actors' social positions are accurate measures of stability by considering the fact that the variability of the latent positions directly affects the variability over time of the probability distribution of the rankings. Thus the stability of the network can be characterized in the proposed model by the precision variables $\tau_t$, $t\geq2$.

\begin{figure}
\centering
\includegraphics[width=0.75\linewidth]{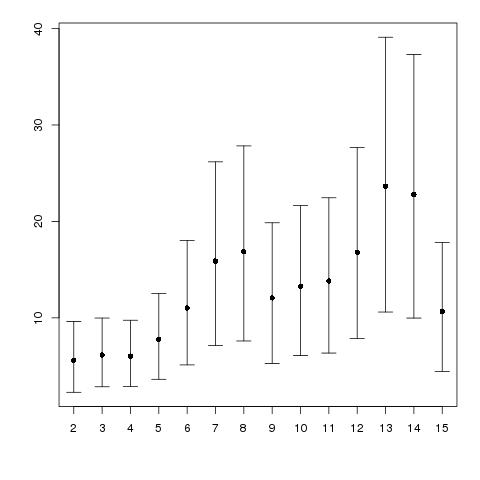}
\caption{Estimates of the precision parameters $\tau_t$, $t=2,\ldots,15$, for the fraternity data.
95\% credible intervals are also given.}
\label{tau_t}
\end{figure}

Our method gives both quantitative point estimates of the network stability as well as uncertainty estimates. Figure \ref{tau_t} gives the posterior means of the $\tau_t$'s and their 95\% credible intervals based on the posterior samples. The higher the precision the more stable the network. The credible intervals in Figure \ref{tau_t} give a good idea as to what values the precision parameters may take, but the intervals cannot be directly compared, i.e., they are not simultaneous credible intervals. Table 1 is given to compare the $\tau_t$'s directly. The $r^{th}$ row $c^{th}$ column entry of this table is the posterior probability that $\tau_{c+1}>\tau_{r+1}$. From Figure \ref{tau_t} we can see the overall pattern of the stability of the network over time, and by using Table 1 we can have more confidence in our inference about the pattern in stability of the network.  For example, looking at Table 1 we see that the posterior probability that $\tau_7>\tau_6$ is 0.85, that $\tau_7>\tau_5$ is 0.96, and that $\tau_7>\tau_2,\tau_3,\tau_4$ each is 0.99, verifying the pattern we see in Figure \ref{tau_t} that the network transitions from week 6 to a more stable form in week 7.

Our results echo that found by using Hoff's multilinear model in that the first few weeks are particularly unstable until around week 6.  Our model also captures the behavior mentioned by Krivitsky and Butts that the network had a downturn of stability heading into the final week of the semester, which is not present in the output of Hoff's model.  We see that even though there is a drastic downturn in network stability, the stability still seems to be above that found in the first five weeks (the probability of the stability being higher in week 15 ranges from 0.75 to 0.91).  This artifact in the data may be due to, as Nakao and Romney suggest, the students becoming distracted during the final week of the semester and of the experiment.

We also detect a new phenomenon in the stability of the network currently unremarked upon by previous analyses of the fraternity data.  From Figure \ref{tau_t} we can see that there is a minor decrease in network stability transitioning from week 8 to week 9.  From Table 1 we see that there is a posterior probability of 0.79 that there is a decrease in stability compared to the previous week, though only a 0.42 probability of having less stability than that observed in week 6 and 0.16 or smaller probability of having less stability than that observed in weeks 1-5.  This is exactly the time when one week of data was not recorded, and one can only conjecture what occurred during this time to decrease the network stability.

The emerging stability within the network implies that the students are making progressively smaller movements over time within the social space.  Looking at Figure \ref{fratLatPos1}, the movements of actors 3, 10, 14, 15 and 16 move progressively towards the edges of the social space, but this is not the same concept as what has been discussed in regards to network stability.  In fact, using our notions of stability, a network could in theory be considered stable while some nodes are moving continually in one direction; in our context we do not in fact see this, but rather most of the actors seem to reach their social position, wherever it may be, and maintain it.

\begin{landscape}
\begin{table}
\caption{Posterior probabilities that $\tau_{c+1}>\tau_{r+1}$ corresponding to the $r^{th}$ row and $c^{th}$ column of the table.}
\label{tau_tTable}
\fbox{
\begin{tabular}{*{15}{r}}
&2&3&4&5&6&7&8&9&10&11&12&13&14&15\\
2&0&0.61&0.58&0.78&0.93&0.99&0.99&0.95&0.97&0.98&0.99 &1.00&1.00&0.91 \\
3&&0&0.49&0.72&0.91&0.99&0.99&0.93&0.96&0.97&0.99 &1.00&1.00&0.88 \\
4&&&0&0.76&0.93&0.99&0.99&0.94&0.97&0.97&0.99 &1.00&1.00&0.89 \\
5&&&&0&0.84&0.96&0.96&0.84&0.89&0.91&0.96 &1.00&0.99&0.75 \\
6&&&&&0&0.85&0.85&0.58&0.67&0.70&0.84&0.96&0.95&0.47 \\
7&&&&&&0&0.58&0.28&0.35&0.38&0.56 &0.82&0.80&0.20 \\
8&&&&&&&0&0.21&0.29&0.33&0.50 &0.78&0.75&0.16\\
9&&&&&&&&0&0.61&0.63&0.78&0.93&0.92&0.39\\
10&&&&&&&&&0&0.55&0.71&0.91&0.89&0.31 \\
11&&&&&&&&&&0&0.71&0.90&0.88&0.28\\
12&&&&&&&&&&&0&0.83&0.77&0.16 \\
13&&&&&&&&&&&&0&0.48&0.04\\
14&&&&&&&&&&&&&0&0.03 \\
15&&&&&&&&&&&&&&0
\end{tabular}
}
\normalsize
\end{table}
\end{landscape}

\subsection{Subgroups}
\label{subgroups}
From early on, researchers have attempted to find well connected subgroups within the overall network; see, e.g., \cite{breiger1975algorithm} and \cite{arabie1978constructing}.  These efforts at what is referred to as community detection were aimed more at demonstrating a new methodology than obtaining any real meaning from the data, making very limited use of the richness in the data.  However, \cite{nakao1993longitudinal} performed a more serious analysis of Newcomb's fraternity data.  The authors embedded Newcomb's fraternity data into a Euclidean space using an ad hoc method of comparing the correlation between actors' rankings and then applying MDS on the resulting similarity matrices; thus two actors would be close together in this space if they ranked the other actors similarly.  Nakao and Romney then used this visualization to determine two subgroups consisting of actors $(1, 5, 6, 8, 13)$ in group one and $(2,4,7,9,11,12,17)$ in group two.  After fitting our model for ranked dynamic networks, we see similar groupings in Figures \ref{fratLatPos1} and \ref{fratLatPos2}.  Nakao and Romney's group one seems to be identically grouped in our visualization, and group two is similarly grouped in our visualization with the exception that actors 4, 9 and 17 seem to form a third, more central, group which bridges group one and group two.  Also, actors 5 and 2 seem to bridge the central group with group one and group two respectively.

The remaining actors, $(3,10,14,15,16)$, were labeled by Nakao and Romney as ``outliers," by which the authors meant that these actors did not find their social positions during the course of the study.  Their visualization has these five actors moving all over the latent space.  However, in our visualization we see that rather than roaming aimlessly, these nodes simply moved farther towards the edge of the social space;  this implies deteriorating friendships rather than allegiance swapping.  \cite{moody2005dynamic} were also able to discover this move towards the edge of the social space in actors 10 and 15 through their visualization methods.

The question remains as to when these subgroups formed.  Nakao and Romney simply state that the subgroups form early in the study and remain stable afterwards.  Using blockmodeling on the binarized network at week 15 to obtain blocks and comparing the proportion of edges between and within blocks at each time point, \cite{arabie1978constructing} claimed that the subgroup formation became stable at week 5.  By partitioning the actors at each time point according to their top four rankings and bottom three rankings and then comparing the partitions over time, \cite{doreian1996brief} claimed that the subgroup formation reached a stable form at week 7.  These methods while all somewhat reasonable are nevertheless rather ad hoc and typically do not make full use of the ranked data.

By using a formal statistical framework to model the fraternity data, we obtain what the other methods do not have: uncertainty estimates.  We utilize these uncertainty estimates to evaluate the timing of the subgroup formation.  From the MCMC output, we can obtain Bayesian credible regions for the latent positions.  If the subgroups have not yet formed we would expect to see these credible regions to be overlapping considerably, i.e., groups of actors are not well separated with high probability, whereas after the subgroups have stabilized we would expect to see overlap in credible regions only in actors belonging to the same subgroup, i.e., low probability that actors of two differing subgroups would be near.

\begin{figure}
\centering
\subfigure[$t=1$]{
\includegraphics[width=0.47\linewidth]{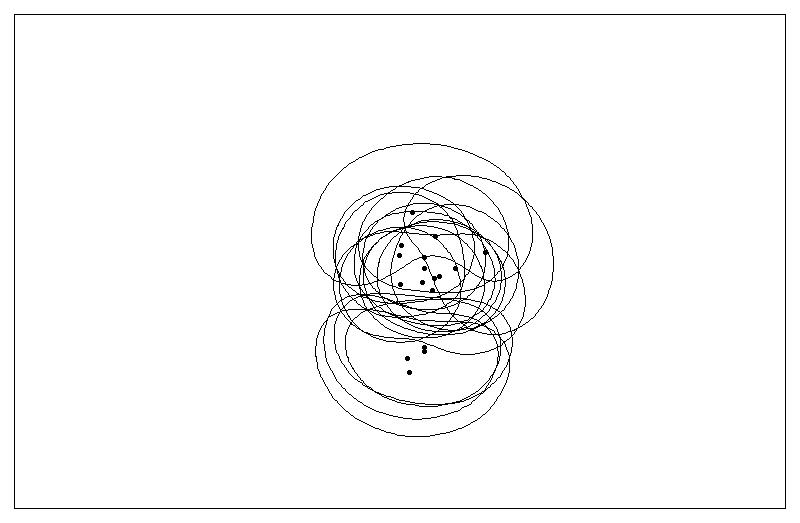}
}
\subfigure[$t=4$]{
\includegraphics[width=0.47\linewidth]{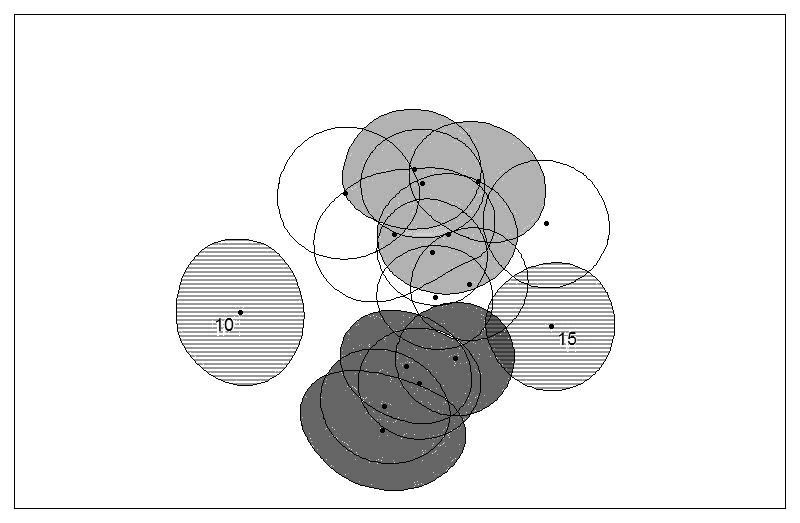}
}\\
\subfigure[$t=6$]{
\includegraphics[width=0.47\linewidth]{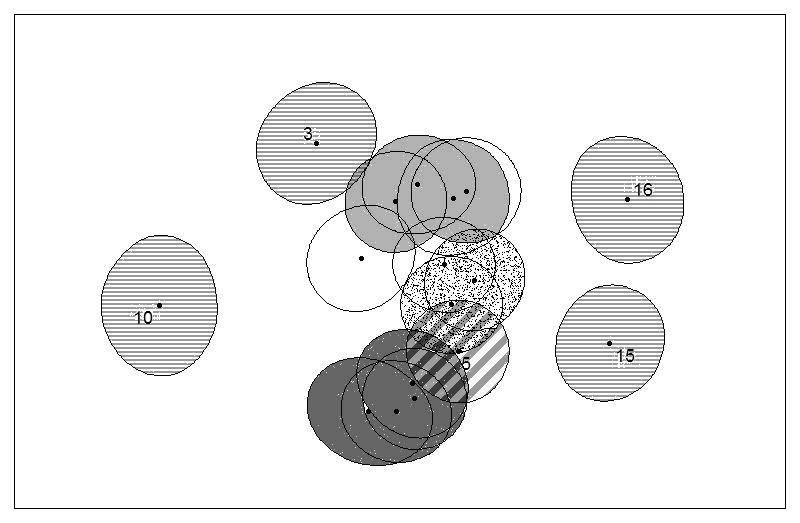}
}
\subfigure[$t=7$]{
\includegraphics[width=0.47\linewidth]{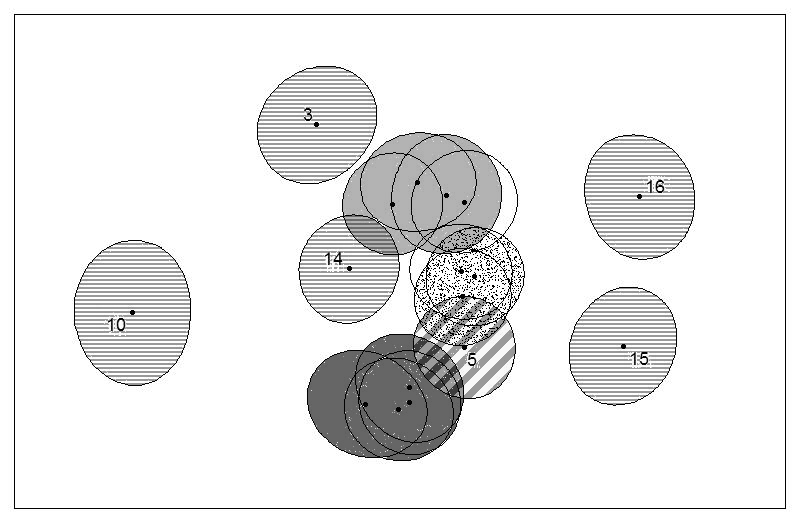}
}\\
\subfigure[$t=9$]{
\includegraphics[width=0.47\linewidth]{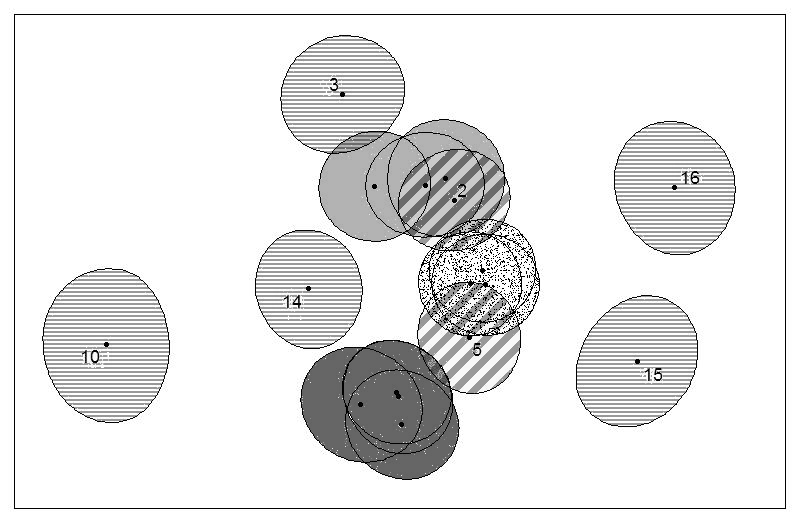}
}
\subfigure[$t=10$]{
\includegraphics[width=0.47\linewidth]{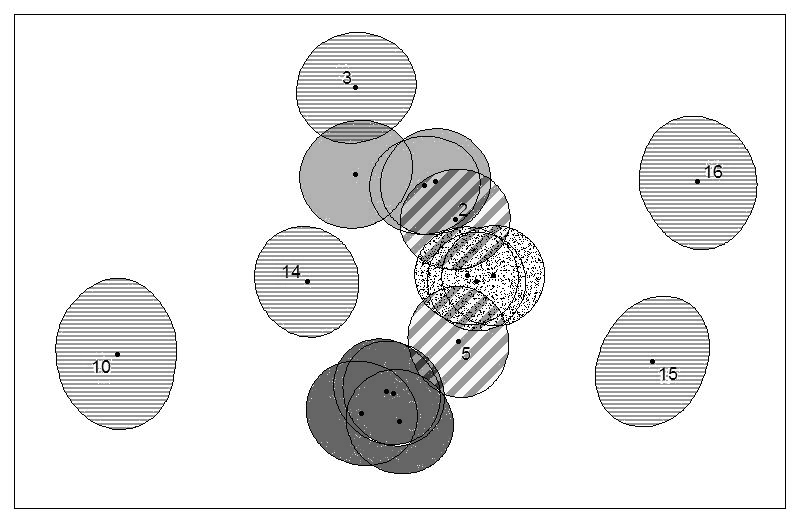}
}
\caption{Plots of 95\% credible regions for latent positions.  The overlap/nonoverlap of the credible regions gives information on the timing of the subgroup formation.  The light, dark and speckled shadings indicate the top, bottom and central subgroups respectively; the horizontal stripes indicate outlying students; diagonal stripes indicate students who bridge two subgroups.}
\label{uncertainty}
\end{figure}

Figure \ref{uncertainty} gives, for $t=1,4,6,7,9,10$, the latent position plots with the 95\% posterior probability regions, using a bivariate density estimation to estimate the boundaries of the regions.  At week 1 we see that there is no subgroup structure at all.  However, by week 4 we see that the top and bottom subgroups have begun to form and are already separated, and also that student 10 and to a lesser degree student 15 are already making their way to the edge of the social space.  At week 6 all three subgroups have started to separate, and at week 7 this structure becomes even more clear.  At week 6 we also see that actor 5 is bridging the bottom and middle subgroups and that actors 3, 10, 15 and 16 have departed from the three main subgroups; at week 7 actor 14 also seems to depart from the three subgroups.  At weeks 9 and 10 the subgroup structure is quite clear, with the final change taking place; this change is due to actor 2 becoming a bridge between the top and middle subgroups.  Although there are some small local changes, it is this structure at week 10 that is in place for the remainder of the study.  We have indicated the top subgroup by a light solid gray shading, the bottom subgroup by a dark solid gray shading, the central subgroup by speckling, the outlying students by horizontal stripes, and the bridging students by diagonal stripes.  Note that at each time point there may be several students who do not belong to any subgroup, in which case there is no shading.

With a small network such as the fraternity data, it was possible to manually determine these subgroups via Figure \ref{uncertainty}.  For larger networks, it should be easier to find these subgroups by considering $n\times n$ adjacency matrices constructed at each time point by setting the $(i,j)$ entry to one if the $i^{th}$ and $j^{th}$ actors have overlapping credible regions.  Thus if two subgroups have separated, we would expect to see blocks of ones along the diagonal corresponding to the closeness of the subgroups and blocks of zeros in the off-diagonals corresponding to the separation between the subgroups.  While we have not experimented with this for larger networks, it should be possible to utilize some standard clustering methods (we were successful applying k-means clustering to the fraternity data) on these adjacency matrices to help find well separated subgroups.  The adjacency matrices constructed from the credible regions of the latent positions of the students in the fraternity data set have been included in the Supplementary Materials, along with a more detailed description of this potential approach to detecting subgroups within larger network data.

It seems reasonable to expect that not all groups would form and stabilize at the same time, and this is what we see here.  The top and bottom groups form first around week 4, the third group forms at week 6 or 7.  Meanwhile, over the first half of the semester certain individuals fail to join a subgroup, moving farther toward the edge of the social space.

\subsection{Popularity and Individual Stability}
We now address the question of whether or not an individual's popularity has any effect on that individual's personal stability within the network.  \cite{nakao1993longitudinal}, by embedding the fraternity data in a latent Euclidean space, claimed that individual stability can be used to predict the individual's position in the final subgroup structure.  In other words, this statement by Nakao and Romney says that actors who have difficulty finding their social position will not find their social position within one of the subgroups.  This is not telling us too much since the subgroup structure was determined by actors which stayed close together in the latent social space, and hence actors that have large movements in the latent space would not tend to stay close to any one particular region of the latent space.  Here we are more interested in discovering whether an individual's popularity, i.e., how well liked an individual is, is related to the individual's stability within the network structure.  That is, does a more popular actor find their social position more effectively than a less popular individual?  \cite{wasserman1980analyzing} used his proposed method to analyze Newcomb's fraternity data to claim that popular individuals remain popular while less popular individuals become even less so over time.  This statement implies some of the movements we see in Figures \ref{fratLatPos1} and \ref{fratLatPos2}, where some individuals stay in the center and others move farther over time towards the edge of the social space.  However, if we take popularity to be an intrinsic time-independent quality of how likeable an individual is, then we still have not answered the question of whether or not popularity is related to individual stability.

Using our proposed model for ranked dynamic networks, we frame our question in terms of finding a relationship between average step size, i.e., $\sum_{t\geq2}\|{\bf X}_{it}-{\bf X}_{i(t-1)}\|/(T-1)$ (we will denote this quantity by $s_i$), with the social reach $r_i$.  A key understanding in this approach is that by including $\br$ in the model, the step sizes are already accounting for the popularity of the individuals.  Hence there is no forced relationship between the step sizes and $\br$ in the model; that is, if $\br$ was not included in the model then an unpopular individual would be forced to move around the outside of the network to maintain low probabilities of receiving favorable rankings, but here that is not the case since we have already accounted for the intrinsic popularity of the individuals.  Therefore any relationship we see between step size and $\br$ is indicative of some fundamental relationship between individual stability and popularity.

\begin{figure}[htb]
\centering
\includegraphics[width=0.6\linewidth]{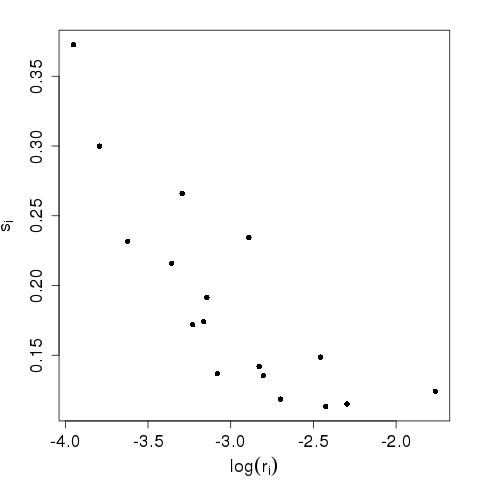}
\caption{Plot of posterior means of the step sizes vs. the log of the posterior means of the social reaches.}
\label{corStER}
\end{figure}

To make sure that $\br$ held the intended meaning of intrinsic likability of an individual, we computed the correlation between the posterior mean of the log of $\br$ with the mean ranking for each individual, averaged over all other nodes at all time points; this correlation was $-0.949$ (recall that a lower ranking is a more favorable ranking), implying that the interpretation of the social reaches is valid.  We then used the posterior means of the latent positions to compute the average step size and the posterior means of the social reaches to estimate the correlation between ${\bf s}=(s_1,\ldots,s_n)$ and $\log(\br)$.  We did comparisons with the $\log(\br)$ because from plotting ${\bf s}$ vs. $\log(\br)$ we see a strong linear relationship (see Figure \ref{corStER}); this is not surprising since the means of $z_{ijt}$ equal $\log(r_j)-d_{ijt}$ (see Section \ref{LSMRDN}), and hence we might have expected to see a linear relationship between the step sizes with the $\log(\br)$.  The correlation was $-0.819$.  Hence we see that there is a strong positive relationship between an individual's intrinsic popularity and the individual's ability to stabilize his social position.

\section{Discussion}
\label{Discussion}
Ranked network data can contain more information than binary network data.  This type of network can be especially important in social networks, such as liking networks or advice-seeking networks.  It is quite possible that with the continuing development of analytical tools for rank-order networks, researchers will collect more meaningful data within this structural framework.

In this work we have proposed a new model for analyzing ranked dynamic networks and used this model to analyze Newcomb's fraternity data.  Our proposed method gives a visualization of the network which allows for a better understanding of its structure and evolution.  Using our proposed model, we investigated  how and when the global network structure stabilizes by incorporating into the model time dependent measures of the network stability.  By using a formal statistical model and estimation procedure we obtained uncertainty estimates of the latent positions which allowed us to evaluate when subgroups formed and stabilized.  Finally, by incorporating individual popularity into the model we ascertained a strong positive relationship between an individual's popularity and individual's stability.

While our model can be applied to any ranked dynamic network for future analyses, one problem that will likely arise is scalability.  Due to the partial sums in the denominator of the likelihood (\ref{PL2}), computing the log likelihood involves summing $O(Tn^3)$ terms, thus rendering most estimation techniques and certainly MCMC methods impractical for large networks.  Newcomb's fraternity data is quite small, and so scalability was not an issue in our analyses, but larger networks may prove too computationally expensive to use our proposed approach, and hence future work in this would be useful for researchers.  For binary static networks, \cite{salter2013variational} developed a variational Bayesian approximation method and \cite{raftery2012fast} used case-control principles to approximate the log likelihood of the latent space model;  this latter approach was further adopted for dynamic binary network data whose likelihood, conditioning on the latent positions, follow an exponential family of distributions \citep{sewell2014weighted}.  These methods would require further work to accommodate our model.  One suggestion from a reviewer was to consider only the top $q$ rankings, thus trading some information for computational feasibility.  This idea was presented by \cite{silverberg1980statistical} as $q$-permutations.  By doing this the computational cost associated with computing the log likelihood would decrease to $O(Tn^2)$.  This should be helpful in medium sized data sets, but further research may still be necessary in developing scalable algorithms for very large networks.

\section*{Acknowledgements}
We thank the Joint Editor, the Associate Editor, and two
referees for their constructive comments which have greatly improved the paper. This work was supported
in part by NSF grants DMS-1106796 and DMS-1406455.

\bibliographystyle{Chicago}

\bibliography{extension}

\end{document}